# From Index to Equity: Pre-Training Transformers for Stock Return Prediction

**Marie Soehl Coolsaet[1], Roberto Gallardo[2], and Zhen Gao[1]**
[1]Faculty of Engineering, McMaster University
[2]Department of Economics, Universidad Veracruzana, Mexico
[*] E-mails: rogallardo@uv.mx; gaozhen@mcmaster.ca

**Abstract:**

This research aims to leverage machine learning to improve stock price prediction and support informed investment decisions related to buying, selling, and holding assets. Specifically, this work investigates transformer-based models for stock prediction and examines the impact of pre-training strategies on forecasting performance. A transformer model was first pre-trained on the Toronto Stock Exchange Index (TSX) to predict intra-day return direction and subsequently fine-tuned on individual TSX stocks. The model was further adapted for return-value regression tasks. Performance was benchmarked against Long Short-Term Memory (LSTM) and XGBoost models. Pre-training on the market index improved the binary cross-entropy loss for individual stock prediction from 0.69 to 0.64. The fine-tuned transformer regression model achieved lower mean squared error than the benchmark models, although the ensemble and XGBoost models achieved higher average daily returns. In addition, a practical application was developed to deliver real-time stock predictions for trading support. Future work will focus on increasing transformer model capacity, incorporating broader global technical indicators, and filtering out stocks with low predictability.

## 1. Introduction

Predicting stock prices is a challenging task because financial markets are influenced by a wide range of economic, political, and social factors. The stock market can be viewed as a chaotic system whose responses vary depending on prevailing conditions and events [1]. Market reactions are often immediate and anomalous, reducing both predictability and response time. In addition, stock market data are inherently non-stationary, requiring forecasting models and algorithms to remain adaptable over time. This non-stationary nature also creates challenges in evaluating predictive models, as the data distribution during testing or hold-out periods may differ significantly from that of the training data. Analyzing stock behavior at an aggregate level, such as through a market index, may help smooth the chaotic fluctuations observed at the individual stock level.

Previous studies in stock prediction have focused on forecasting short-term price direction as well as short- and long-term returns. Common input features include sequences of recent stock prices, historical performance metrics, volatility measures, and broader economic indicators. Traditional machine learning approaches for stock prediction include support vector machines (SVM), tree-based and ensemble methods, recurrent neural networks (RNN), and Long Short-Term Memory (LSTM) networks. More recently, transformer architectures have been applied to financial forecasting and have demonstrated strong performance relative to existing approaches [2].

Transformers, first introduced in 2017 [3], revolutionized the field of natural language processing and now form the foundation of modern large language models, including ChatGPT [4]. These models have demonstrated remarkable capabilities in question answering, code generation, and a wide range of natural language tasks. The success of transformers is largely attributed to the multi-head self-attention mechanism [3]. Self-attention computes relationships between input elements by evaluating their interactions through dot-product operations, enabling the model to generate weighted feature representations based on contextual relevance [3]. The multi-head design further enhances model flexibility by allowing multiple attention representations to be learned simultaneously [3].

This mechanism has proven highly effective for analyzing sequential data, such as sentences and paragraphs, by dynamically assigning importance to input features based on the current context or task. In the context of stock market prediction, transformers may similarly contextualize stock behavior using technical and financial indicators as input features. The transformer architecture consists of an encoder and a decoder [3]. The encoder generates learned representations of the input sequence, while the decoder iteratively predicts subsequent outputs based on previously generated terms [3]. This architecture enables transformers to generate sequential outputs, such as text or code. In financial forecasting applications, transformers can be adapted to predict stock prices across multiple future trading days. Alternatively, by using only the encoder component, the model can be configured to produce a single prediction value, such as the stock price or return for the following trading day.

For transformers to effectively learn contextual relationships between words and sentences, they typically require training on extremely large datasets, often involving months of computationally intensive training. Historically, a major limitation in natural language processing tasks was the availability of sufficiently large and well-labeled datasets, such as question–answer pairs for supervised learning. Large language models addressed this challenge through the use of pre-training tasks, enabling models to learn general linguistic structures and semantic relationships before being fine-tuned for downstream applications [5]. Common pre-training tasks include masked language modeling, where the model predicts missing words within a sentence, and next-sentence prediction, where the model determines the most likely subsequent sentence given an input sequence.

## 2. Method

The goal of this study was to improve the prediction accuracy of a transformer model for daily stock returns on the Toronto Stock Exchange by using pre-training techniques. It utilized the transformer architecture for stock price prediction. Stock market data were accessed through the Yahoo Finance API1, and relevant features were engineered as model inputs. Transformer models, LSTM models, an XGBoost gradient boosting model, and an ensemble model were trained to predict intra-day price returns. In addition, an application was developed to generate predictions for current trading dates and visualize financial indicators.

The dataset focused on the Canadian stock market, specifically stocks listed on the Toronto Stock Exchange (TSX). Data were retrieved through the Yahoo Finance API using the yfinance Python package. A list of TSX ticker symbols was downloaded directly from Yahoo Finance and initially included 2,310 tickers. For each ticker, historical open, close, high, and low prices, along with trading volume, were collected for all available dates. Because data for some stocks were unavailable through the API, the final dataset contained 1,853 tickers. The earliest available record dated back to March 17, 1980, while the latest date used was July 4, 2023. In total, the dataset consisted of 4,915,504 daily stock records spanning 15,814 trading dates. Figure 1 presents a histogram of the number of records by date, illustrating that more recent periods contain substantially greater data availability.

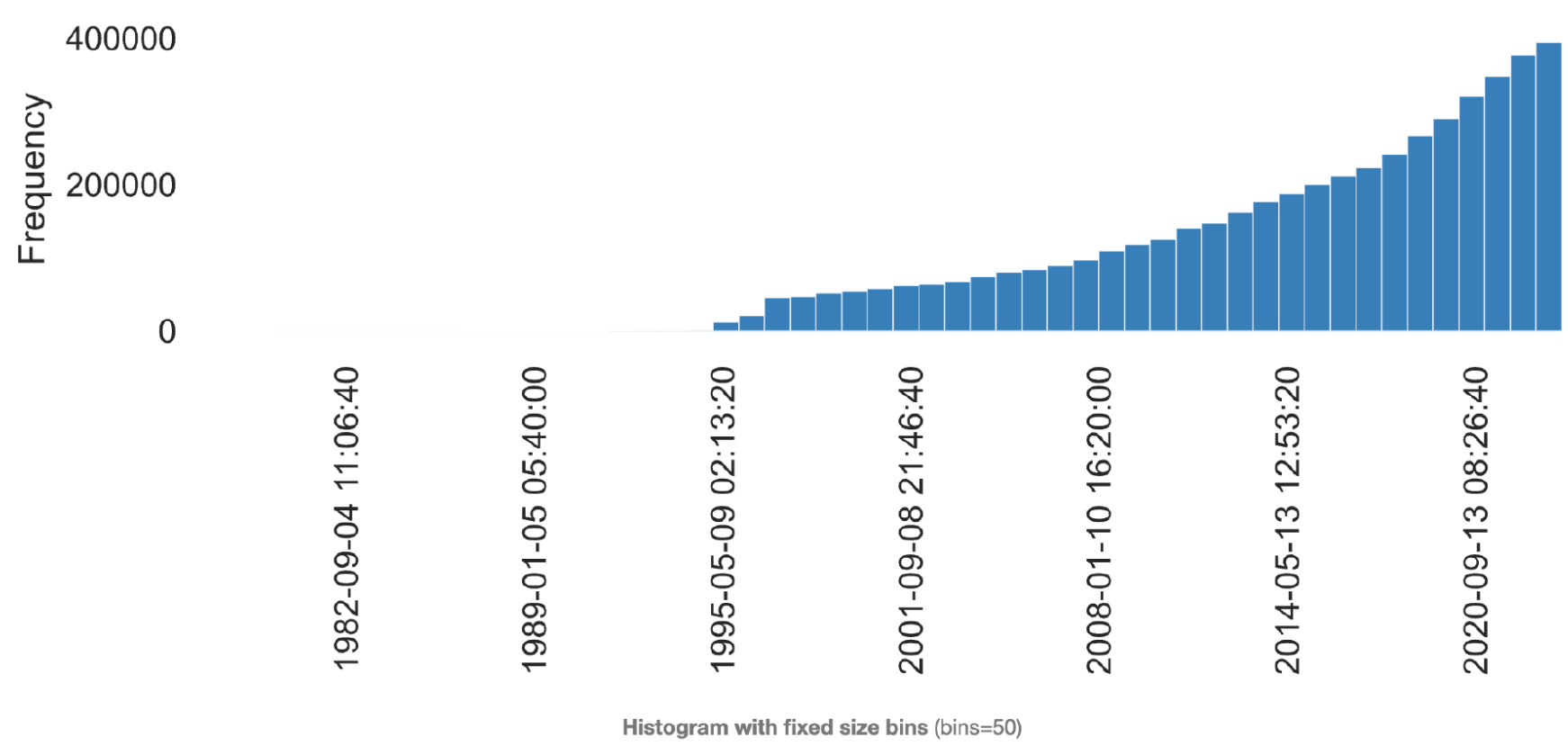

Figure 1. Count of daily stock performance records by date

Features were engineered to capture the historical behavior of stocks and various technical financial indicators. The feature engineering process closely followed the methodology presented in [6]. The intraday price return for the prediction date was used as the model target. Because stock prices tend to increase over time, the use of intra-day price returns provides a standardized measure of daily performance, enabling meaningful comparisons across stocks with different price ranges.

In addition to historic lag features, various technical indicators were used as features [6]. These indicators are widely used in the financial industry to provide insight into stock price trends and market behavior. Stochastic %K is a stochastic oscillator in which increasing values generally indicate that a stock price is likely to rise, while decreasing values may signal a potential decline. Momentum-based indicators include the price rate of change (ROC), disparity 5, and disparity 10. ROC can help identify overbought or oversold conditions in the market. A positive disparity value suggests an upward price trend with increasing momentum, whereas a negative disparity value may indicate a likely price decline. The relative strength index (RSI) is another commonly used momentum indicator for identifying overbought or oversold conditions; specifically, RSI values above 70 may indicate that a stock is overbought and likely to decline, while values below 30 may suggest that the stock is oversold and likely to increase in price. Finally, trend-following indicators, including the accumulation distribution oscillator (AccDO), moving average convergence-divergence (MACD), and exponential moving average (EMA), are used to track market trends and assess the strength of those trends.

These indicators, along with the intra-day price return, day of the year, and additional metrics used in indicator calculation, such as moving averages, the closing price from 10 days prior, and exponential moving averages, were computed for each stock and trading date in the dataset. Features from the current trading day and the previous nine days were used as model inputs. For the transformer and LSTM models, the input data were structured as a sequence of 10 days with 15 features per day. For the XGBoost model, the sequential inputs were flattened into a vector containing 150 features. The 15 input features used for each of the 10 days are listed here: intra-day price return, exponential moving averages (EMA) for 10, 12, and 26 days, stochastic %K, price rate of change (ROC), relative strength index (RSI), accumulation distribution oscillator (AccDO), moving average convergence-divergence (MACD), disparity indicators for 5-day and 10-day periods, moving averages for 5-day and 10-day periods, and closing price from 10 days prior.

For the classification task, the model target was assigned a value of 1 when the intra-day return was greater than zero and 0 otherwise. Class weighting was applied to address class imbalance in the dataset. For the regression task, the target variable was the intra-day return itself. Both the input features and target values were normalized using min-max scaling, with the minimum and maximum values determined from the training dataset. Training data consisted of stock market records from 2010 to 2022. Validation data included stock data from 2022, while the hold-out test dataset consisted of data from 2023 up to December 1, 2023. The validation dataset was used to assess model performance and guide design decisions related to feature engineering and model architecture. Final model evaluation was performed using the test dataset.

Several techniques were applied to handle outliers and improve data quality. Infinite values occurred in the technical indicators %K and AccDO when the highest high was equal to the lowest low, resulting in division-by-zero errors. For %K, infinite values were replaced with 50, while infinite AccDO values were replaced with 0. These replacement values were selected because they represent neutral midpoint thresholds for the respective indicators. In addition, a small number of significant outliers (18 observations) were identified in the intra-day return values within the training dataset. These outliers disproportionately affected min-max scaling by increasing the maximum value, causing the majority of scaled values to become very small. Because removing the outliers would also eliminate multiple sequential training samples, the outlier values were clipped by setting all intra-day returns greater than 2 equal to 2. To further reduce unpredictable stock behavior, the first year of available records for each stock was excluded from the analysis.

The transformer architecture consisted of four transformer encoder blocks followed by three fully connected dense layers with 128, 64, and 32 nodes, respectively. All dense layers used the ReLU activation function, and positional encoding was applied to the input sequences. For the LSTM benchmark models, the transformer encoder blocks were replaced with four LSTM layers. An overview of the model architecture is presented in Figure 2.

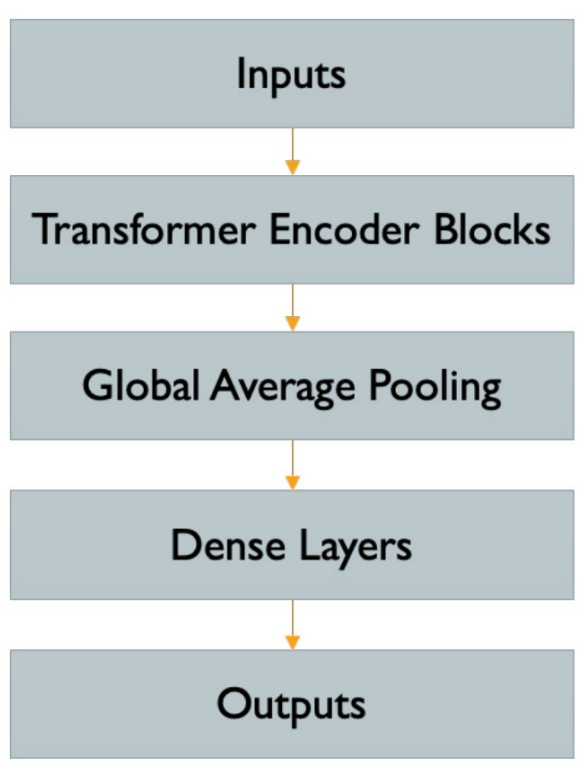

Figure 2. Layers in transformer encoder architecture

For the classification task, the output layer consisted of a dense layer with a single node and a sigmoid activation function. For the regression task, the output layer also consisted of a single-node dense layer but used a linear activation function. Binary cross-entropy was used as the loss function for classification, while mean squared error (MSE) was used for regression. The XGBoost model was trained with the parameters defined in table 1.

Table 1: Parameters used for training the XGBoost model

| **Parameter** | **Value** |
|---|---|
| Number of Estimators | 100 |
| Learning Rate | 0.03 |
| Maximum Depth | 9 |
| Column Sample by Tree | 0.7 |
| Maximum Leaves | 100 |

## 3. Experiment

The first experiment compared two transformer-based approaches for individual stock prediction: one trained with randomly initialized weights and another initialized using weights from a model pre-trained on the market index. For the pre-training stage, historical index data underwent the same preprocessing and feature engineering procedures as the individual stock data. Both models were trained to classify the intra-day return direction as either positive or negative. Classification performance was benchmarked against LSTM and XGBoost models. Similar to the transformer approach, the LSTM model was first pre-trained on the market index and then fine-tuned using individual stock data. An additional benchmark consisted of an ensemble model combining the fine-tuned deep learning models with the XGBoost model. Ensemble predictions were generated by averaging the predicted outputs of the individual models.

The second experiment evaluated regression performance. The transformer and LSTM classification models were further fine-tuned for regression tasks by replacing the output layer. These models were trained to predict the intra-day return values of individual stocks. A regression-based XGBoost model and an ensemble regression model were also used as benchmarks. Following the methodology described in [6], stocks in the test dataset were ranked according to their predicted intra-day returns. For each trading day, the five stocks with the highest predicted returns were selected for purchase, while the five stocks with the lowest predicted returns were shorted. Positions were held for a single trading day, and the average return over the test period was subsequently calculated.

A Python-based application was developed using Streamlit to provide an interactive interface for applying forecasted stock performance to trading decisions. The application integrates multiple components of this research, including data acquisition, feature engineering, and model inference. It retrieves the latest market data through the Yahoo Finance API and generates the required model input features. Predictions from the three regression models are obtained and averaged to produce the final forecast. Stocks are then ranked according to their predicted returns, and the application displays the top k and bottom k stocks, where k is specified by the user. Figure 3 illustrates the prediction interface for the next available trading date within the application.

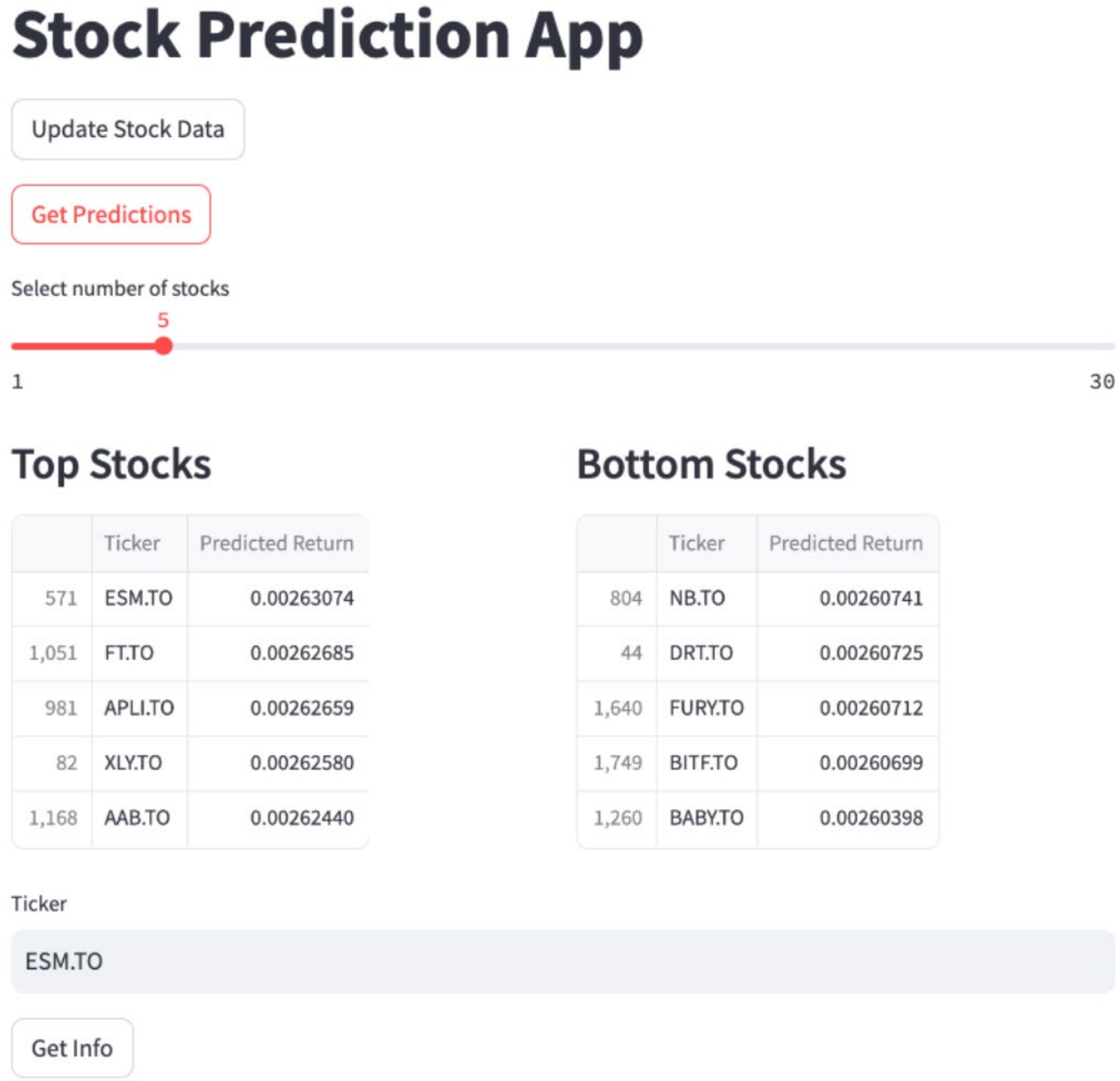

Figure 3. A screen in the application that shows a button to update stock data and get predictions

A search bar was added to the application, allowing users to enter a stock ticker and view recent trends in the corresponding technical indicators. The search interface is shown in Figure 4.

Figure 4. A screen in the application that has search input to specify a stock and provides information about historical performance and technical indicators

## 4. Results and Discussion

The results for the classification of daily intraday stock return direction (increase or decrease) are summarized in Table 2. When the baseline transformer model was trained on individual stocks using randomly initialized weights, performance failed to improve beyond the first training epoch, resulting in a binary cross-entropy loss of 0.6942. In contrast, pre-training the transformer on the TSX index followed by fine-tuning on individual stock data reduced the loss to 0.6404. These results demonstrate the effectiveness of pre-training on aggregated market data, which enables the model to learn more generalized patterns from less chaotic data. The aggregation process helps smooth the highly volatile behavior observed at the individual stock level.

Table 2: Classification results for individual stock prediction

| | **Transformer Baseline** | **Transformer Fine-Tuned** | **LSTM Fine-Tuned** | **XGBoost** | **Ensemble** |
|---|---|---|---|---|---|
| Test Accuracy | 0.6667 | 0.6572 | 0.6660 | 0.6272 | 0.6594 |
| Validation Accuracy | 0.6634 | 0.6514 | 0.6624 | 0.6109 | 0.6542 |
| Train Accuracy | 0.6343 | 0.6183 | 0.6329 | 0.5776 | 0.6217 |

| Test Precision | 0.0000 | 0.3326 | 0.3840 | 0.4626 | 0.3434 |
|---|---|---|---|---|---|
| Validation Precision | 0.0000 | 0.3589 | 0.3184 | 0.4540 | 0.3734 |
| Train Precision | 0.0000 | 0.3310 | 0.4057 | 0.4555 | 0.3433 |
| Test F1 Score | 0.0000 | 0.0523 | 0.0072 | 0.5675 | 0.0451 |
| Validation F1 Score | 0.0000 | 0.0804 | 0.0049 | 0.5697 | 0.0727 |
| Train F1 Score | 0.0000 | 0.0760 | 0.0169 | 0.5789 | 0.0683 |

Although the baseline transformer model achieved the highest test accuracy, examination of the precision and F1 score indicates that the model effectively predicted only a single class. Due to the class imbalance in the dataset, where negative stock return days occurred more frequently than positive return days, the F1 score serves as a more reliable indicator of classification performance. The fine-tuned transformer model outperformed the fine-tuned LSTM model; however, the XGBoost model achieved the strongest overall classification performance.

The regression models were trained to predict standardized intra-day returns for individual stocks. The transformer and LSTM regression models were initialized using the previously trained classification models and subsequently fine-tuned for regression tasks. Both models outperformed the XGBoost regression model, further demonstrating the benefit of pre-training on aggregated market-level data before fine-tuning on more volatile and chaotic individual stock data. In this case, the pre-training task involved learning an aggregated target, i.e., binary return direction, before adapting the models to predict the more complex continuous return values. The regression results are summarized in Table 3.

Table 3. Regression results for individual stock prediction

| | **Transformer Fine-Tuned** | **LSTM Fine-Tuned** | **XGBoost** | **Ensemble** |
|---|---|---|---|---|
| Test MSE | 5.682e-9 | 5.683e-9 | 9.923e-9 | 6.147e-9 |
| Validation MSE | 6.955e-9 | 6.961e-9 | 3.300e-8 | 9.844e-9 |
| Train MSE | 4.477e-7 | 4.478e-7 | 5.386e-8 | 2.640e-7 |

The fine-tuned transformer model achieved a test mean squared error (MSE) of 5.682e-9, slightly outperforming the fine-tuned LSTM model, which achieved a test MSE of 5.683e-9. Both fine-tuned deep learning models outperformed the XGBoost and ensemble models in terms of regression accuracy. All models achieved positive average returns over the test period. Among them, the ensemble model produced the highest average return. The portfolio return results are summarized in Table 4.

Table 4. Average returns for daily portfolio rebalancing with the top 5 and bottom 5 stocks

| **Model** | **Average Return** |
|---|---|
| Transformer Fine-Tuned | 0.0003 |
| LSTM Fine-Tuned | 0.0002 |
| XGBoost | 0.0059 |

| Ensemble | 0.0069 |
|---|---|

Interestingly, the model achieving the highest average return did not correspond to the model with the best mean squared error performance. One possible explanation is that the stocks selected by the transformer and LSTM models tended to have predicted returns closer to zero, whereas the stocks selected by the XGBoost model exhibited greater return variation. This behavior is illustrated in Figure 5, which presents the weekly average returns generated by each model.

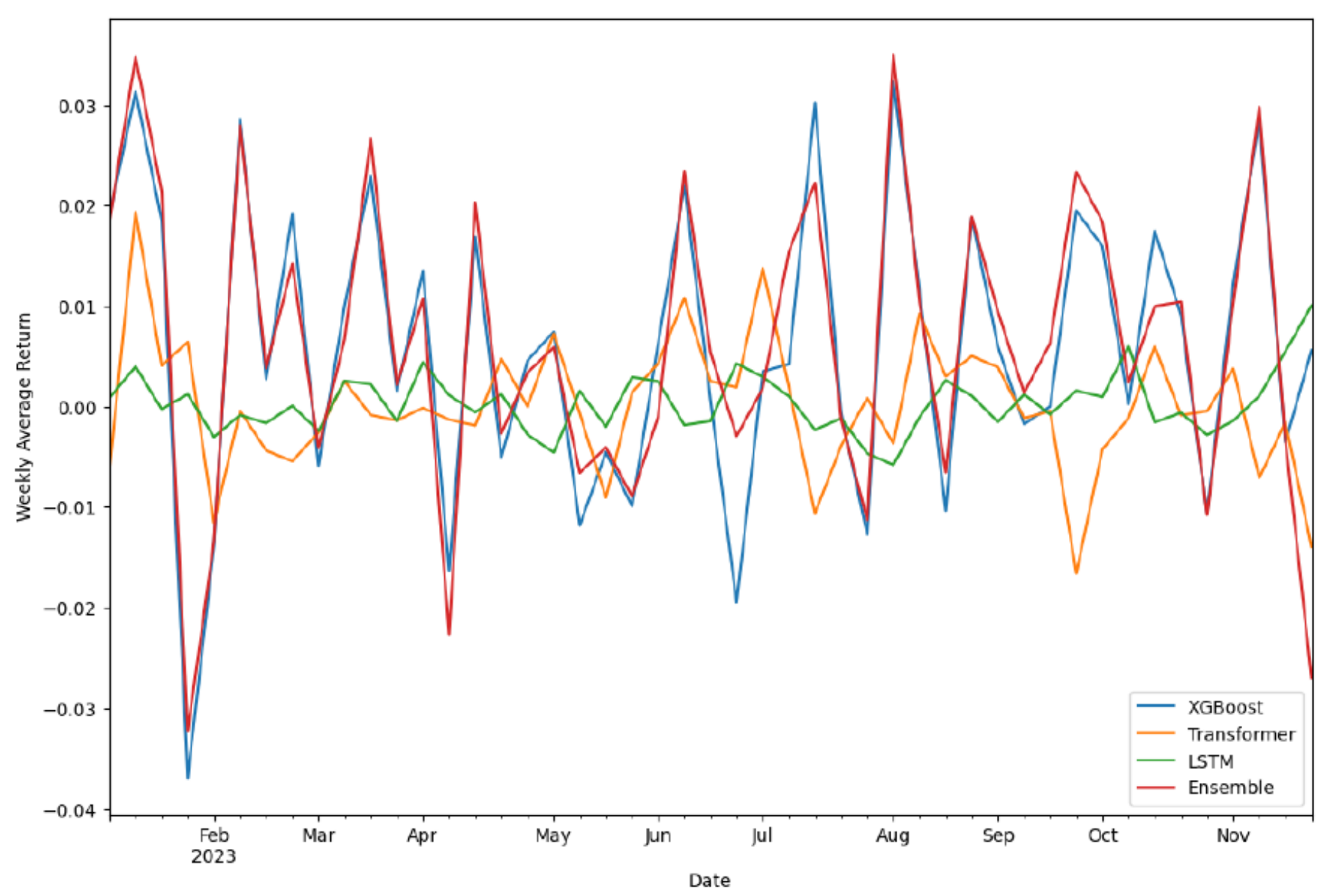

Figure 5. Weekly average returns of individual stocks

## 5. Conclusions and Future Work

The primary objective of this research was to investigate the use of transformer models for individual stock market prediction and evaluate the impact of pre-training techniques on forecasting performance. Although the average returns achieved during the test period were positive, they remained relatively modest. These results, together with the classification and regression performance across all models, highlight the inherent difficulty of working with stock market data. Nevertheless, pre-training on the market index demonstrated promising potential for improving individual stock prediction performance. In addition, transfer learning from classification to regression proved beneficial, as both the transformer and LSTM models outperformed the XGBoost model in the regression task.

Future work will focus on increasing the capacity of the transformer architecture to capture more complex patterns in the data. In addition, broader global financial indicators will be incorporated into the input features, as the current implementation relies primarily on historical performance data from individual stocks. The proposed methodology could also be extended to other market indexes and may achieve stronger performance on markets with higher trading volume and more stable trading dynamics than the Toronto Stock Exchange. This project also did not explicitly consider portfolio risk management in stock recommendations. A risk-control strategy proposed in [6] involved excluding stocks with high historical prediction error over the most recent 40-day period from portfolio selection, and this approach could be incorporated into the application layer. Finally, to leverage recent advances in large language models (LLMs), future research could explore transformer architectures that incorporate pre-trained components derived from LLMs.

**References**

1. Z. Liu, "Chaotic Time Series Analysis," Mathematical Problems in Engineering, vol. doi:10.1155/2010/720190, 2010.
2. D. Kisiel and D. Gorse, "Portfolio Transformer for Attention-Based Asset Allocation," arXiv:2206.03246v1, 2022.
3. A. Vaswani, N. Shazeer, N. Parmar, J. Uskoreit, L. Jones, A. N. Gomez, L. Kaiser and I. Polosukhin, "Attention Is All You Need," arXiv:1706.03762v5, 2017.
4. OpenAI, "ChatGPT," 2021. Available: https://openai.com/research/chatgpt.
5. W.-C. Chang, F. X. Yu, Y.-W. Chang, Y. Yang and S. Kumar, "Pre-training tasks for embedding-based large-scale retrieval," arXiv:2002.03932v1, 2020.
6. M. S. Carta, S. Consoli, A. S. Podda, D. R. Recupero and M. M. Stanciu, "Ensembling and Dynamic Asset Selection for Risk-Controlled Statistical Arbitrage," IEEE Access, vol. 9, pp. 29942-29959, 2021.
7. T. Zhou, P. Niu, X. Wang, L. Sun and R. Jin, "One Fits All: Power General Time Series Analysis by Pretrained LM," arXiv:2302.11939v4, 2023.
8. Anonymous authors, "[Under double blind review] PromptCast: A new prompt-based learning paradigm for time series forecasting," International Conference on Learning Representations, 2023.